\documentclass[a4paper,11pt]{article}
\usepackage{jheppub} 
\usepackage{lineno}
\usepackage{graphicx}
\usepackage{tikz}
\usepackage{amsbsy,amsmath,amssymb}
\usepackage{dsfont}
\usepackage{amsmath}
\usepackage{amsfonts}
\usepackage{bbold}
\usepackage{filecontents}
\usepackage{mathtools}
\usepackage{empheq}
\usepackage{slashed}
\usepackage{bbm}
\usepackage{microtype}
\usepackage[macrosonly]{chet}
\usepackage{xcolor}
\definecolor{darkblue}{rgb}{0.1,0.1,.7}
\definecolor{darkgreen}{rgb}{0,.5,0}
\usepackage{hyperref}
\usepackage{booktabs}
\usepackage{array}
\usepackage{enumitem}
\usepackage{comment}
\usepackage{mdframed} 
\newmdenv[linecolor=black,linewidth=1pt,roundcorner=6pt, backgroundcolor=gray!10,skipabove=10pt,skipbelow=10pt,innerleftmargin=10pt,innerrightmargin=10pt, innertopmargin=8pt, innerbottommargin=8pt]{mybox}

\newcommand{\cO}{{\cal O}}

\DeclareMathOperator{\Tr}{Tr}

\renewcommand{\a}{\alpha}
\renewcommand{\b}{\beta}

\newcommand{\m}{\mu}
\newcommand{\n}{\nu}

\def\be{\begin{equation}}
\def\ee{\end{equation}}
\def\bea{\begin{eqnarray}}
\def\eea{\end{eqnarray}}
\def\ba{\begin{array}}
\def\ea{\end{array}}
\def\bi{\begin{itemize}}
\def\ei{\end{itemize}}

\def\Tr{{\rm Tr}}

\newcommand{\beq}{\begin{equation}}
\newcommand{\eeq}{\end{equation}}
\newcommand{\beqn}{\begin{eqnarray}}
\newcommand{\eeqn}{\end{eqnarray}}
\newcommand{\bga}{\begin{align}}

\def\dalemb#1#2{{\vbox{\hrule height .#2pt
\hbox{\vrule width.#2pt height#1pt \kern#1pt
\vrule width.#2pt}
\hrule height.#2pt}}}
\def\square{\mathord{\dalemb{6.8}{7}\hbox{\hskip1pt}}}

\def\a{\alpha}
\def\b{\beta}

\def\sc{{\hat{c}}}

\def\cZ{{\cal Z}}

 \def\IZ{\relax\ifmmode\mathchoice
 {\hbox{\cmss Z\kern-.4em Z}}{\hbox{\cmss Z\kern-.4em Z}}
 {\lower.9pt\hbox{\cmsss Z\kern-.4em Z}}
 {\lower1.2pt\hbox{\cmsss Z\kern-.4em Z}}\else{\cmss Z\kern-.4em Z}\fi}
 \def\IB{\relax{\rm I\kern-.18em B}}
 \def\IC{{\relax\hbox{$\inbar\kern-.3em{\rm C}$}}}
 \def\Ic{{\relax\hbox{$\inbar\kern-.22em{\rm c}$}}}
 \def\ID{\relax{\rm I\kern-.18em D}}
 \def\IE{\relax{\rm I\kern-.18em E}}
 \def\IF{\relax{\rm I\kern-.18em F}}
 \def\IG{\relax\hbox{$\inbar\kern-.3em{\rm G}$}}
 \def\IGa{\relax\hbox{${\rm I}\kern-.18em\Gamma$}}
 \def\IH{\relax{\rm I\kern-.18em H}}
 \def\II{\relax{\rm I\kern-.18em I}}
 \def\IK{\relax{\rm I\kern-.18em K}}
 \def\IP{\relax{\rm I\kern-.18em P}}

\def\Tr{{\rm Tr}}
 \font\cmss=cmss10 \font\cmsss=cmss10 at 7pt
 \def\IR{\relax{\rm I\kern-.18em R}}

\def\cO{{\cal O}}

\def\a{\alpha}
\def\b{\beta}

\def\m{\mu}
\def\n{\nu}

\def\cC{{\cal C}}
\def\cQ{{\cal Q}}

\def\cL{{\cal L}}
\def\cF{{\cal F}}

\title{\boldmath Dimensional uplifts of thermal two-point functions and higher-derivative CFTs}

\author{Songyuan Li$^{a,b}$ and Anastasios C. Petkou$^b$} 

\affiliation{$^a$School of Physics, Sun Yat-sen University, \\Guangzhou 510275, China}
\affiliation{$^b$Laboratory of Theoretical Physics,  School of Physics, \\Aristotle University of Thessaloniki, \\54124 Thessaloniki, Greece}

\emailAdd{lisy353@mail.sysu.edu.cn}
\emailAdd{petkou@auth.gr}

\abstract{
We explore the idea that the mass parameter $m$ can be used as a diagnostic tool for identifying the critical points of thermal two-point functions of massive free scalar fields in odd $d$ dimensions. For general values of $m$ these correlators admit an expansion in thermal conformal blocks associated with symmetric traceless operators of definite scaling dimension. For each spin the higher-twist contributions can be resummed into a closed-form dressing factor of the leading-twist coefficient. At special values of $m$ for which the thermal trace of the energy-momentum tensor vanishes they are identified as two-point functions of thermal CFTs. Through a dimensional uplift, valid for all $m$, we map these correlators to thermal two-point functions of higher-derivative massive theories in $d+2$ dimensions. We show that the thermal trace of the energy-momentum tensor of the $d+2$-dimensional theory is proportional to the free energy of the original $d$-dimensional theory.  When that thermal trace vanishes for $m\neq 0$ the uplifted correlators are identified as two-point functions of a higher-derivative nonunitary thermal CFT$_{d+2}$, but they differ from the uplifted nontrivial thermal two-point functions of the original unitary CFT$_d$.

\vspace{2cm}
\begin{center}
\emph{Dedicated to the memory of Robert G. Leigh, who left us too soon, yet whose enduring legacy continues to inspire our pursuit of the fundamentals.}
\end{center}
}

\begin{document}

\maketitle

\section{Introduction}
\label{sec:intro}

Conformal field theories (CFTs) represent anchor points in the ocean of quantum field theories (QFTs).  Identifying and understanding them is expected to provide a navigation map of renormalization-group (RG) flows across the space of QFTs.  CFTs are also of considerable interest in their own right due to their wide range of physical applications, from the description of quantum criticality at zero temperature to the study of strongly correlated systems and universal behavior at finite temperature (see, e.g. \cite{Cardy:1989da,DiFrancesco:1997nk,Sachdev:2011fcc}).

Although the study of CFTs at finite temperature, the so-called thermal CFTs,  has a long history,\footnote{See, e.g. \cite{Petkou:1998wd,Petkou:1998fb,Christiansen:1999uv} and references therein for earlier works.} it has recently attracted renewed interest driven mainly by efforts to understand holographically black hole physics 
\cite{Karlsson:2021duj,Rodriguez-Gomez:2021pfh,Rodriguez-Gomez:2021mkk,Karlsson:2022osn,Dodelson:2022yvn,Dodelson:2023vrw,Ceplak:2024bja}, and also by attempts to extend bootstrap methods to finite temperature \cite{Iliesiu:2018fao,Petkou:2018ynm,Iliesiu:2018zlz,Barrat:2025nvu,Buric:2025anb,Buric:2025fye}. This has led to a wealth of analytic results for correlators and one-point functions \cite{Georgiou:2022ekc,Bhatta:2022wga,Marchetto:2023fcw,Diatlyk:2023msc,Marchetto:2023xap,Alkalaev:2024jxh,Buric:2024kxo,Buric:2025uqt,Komargodski:2026ain,Alkalaev:2026sha,Dodelson:2025jff,Dodelson:2025rng,Barrat:2025twb,Arnaudo:2026der,Buric:2026pes,Araya:2026shz,Arnaudo:2026axe,Barrat:2026jfg,Giombi:2026kdz}, for vector models \cite{David:2023uya,David:2024naf,Barrat:2025wbi}, for novel computational tools \cite{,Niarchos:2025cdg,Barrat:2024aoa,Ghosh:2026jbw,Ghosh:2026xnp,Niarchos:2026vow,Ghosh:2026efw} and for effective actions \cite{Bobev:2023ggk,Parmentier:2026aqh,David:2026yis,Advant:2026cqt}.

The thermal state at temperature $\beta =1/T$ can be introduced in a QFT$_d$ by placing the latter  in the non-trivial geometry $S^1_\b\times\mathbb{R}^{d-1}$ and imposing appropriate boundary conditions along the thermal circle - periodic for bosons and antiperiodic for fermions. Although the thermal state breaks conformal invariance one might expect that the finite temperature CFTs can be determined by their zero temperature properties.  This is the case in $d=2$ \cite{Cardy:1989da}. However, thermal CFTs in $d>2$ seem to contain information that is not easily, if at all, associated with CFT data at $T=0$. The extra information comes in the form of nonzero thermal one-point functions of quasiprimary operators. The latter constitute the coefficients of the expansion of  two-point functions in terms of thermal conformal blocks. 

It is reasonable to expect that RG flows among zero temperature CFTs have a reflection in the space of CFTs at finite temperature \cite{Appelquist:1999hr,Klebanov:2011gs}. Reversing this logic one might try to use thermal two-point functions as diagnostic tools since their critical points can be associated with thermal CFTs  which in turn provide information about the corresponding CFTs at zero temperature. In this work we elaborate on this idea using results from our earlier work on thermal two-point functions of massive free scalar theories in odd $d$-dimensions \cite{Petkou:2018ynm,Petkou:2021zhg,Karydas:2023ufs,Karydas:2025tfs}.  We show that for fixed temperature the mass parameter $m$ explores the theory space near massless thermal CFTs and allows us to identify nontrivial ones when a thermal gap equation is satisfied. The main technical tool is the expansion, for general values of $m$,  of the two-point functions in thermal conformal blocks that are associated with contributions from higher-spin symmetric traceless tensor operators with definite scaling dimensions. At leading-twist these have dimensions $\Delta^{(d)}_s=d-2+s$ that saturate the unitarity bound and turn into higher-spin conserved currents at the critical points when the correlator is associated with a thermal CFT.  We show that the contributions from higher-twist operators, for each spin, can be resummed into dressing factors of the leading-twist coefficients in the expansion. Assuming that the theory is local and hence it possesses an energy-momentum tensor we obtain  the thermal gap equation that needs to be satisfied in order that the two-point function is associated with a thermal CFT$_d$. The gap equation is identified as the thermal trace of the energy-momentum tensor of the theory.

We then observe that $d$-dimensional thermal conformal blocks can be straightforwardly uplifted to a linear combination of thermal conformal blocks in $d+2$-dimensions
associated with contributions from two distinct classes of operators: I) Higher-spin symmetric traceless tensors with dimensions $d+s$  that saturate the unitarity bound and correspond to the would-be conserved currents of the uplifted theory. II) Higher-spin symmetric traceless tensors with dimensions $d-2+s$ that lie below the unitarity bound. It indicates that the uplifted correlator belongs to a $d+2$-dimensional higher-derivative theory which we identify. The thermal one-point function of the energy-momentum tensor in the $d+2$-dimensional nonunitary theory is related to the spin-4 one-point function of the original $d$-dimensional unitary theory. As a consequence, the thermal gap equation of the uplifted theory differs from that of the original theory, and hence the uplifted thermal two-point functions of the corresponding CFT$_d$ are not conformal thermal two-point functions of the uplifted nonunitary CFT$_{d+2}$ when the thermal mass is nonzero.  Notably, the gap equation of the uplifted correlator is identified with the free energy of the original unitary $d$-dimensional theory. 

In Section \ref{sec:2ptfunctioninddim} we present several analytic results regarding thermal two-point functions of massive free scalars. In Section \ref{sec:gapeq} we discuss the thermal gap equation. In Section \ref{sec:uplift} we present the thermal dimensional uplift and the higher-derivative theory that gives rise to the uplifted thermal two-point function. Possible consequences and extensions of our results are discussed in Section \ref{sec:discussion}. The Appendices contain technical details of the calculations in the main text. 

\section{Thermal two-point functions of scalar fields in $d$-dimensions}\label{sec:2ptfunctioninddim}

\subsection{Thermal two-point functions of scalar CFTs }

The thermal two-point function of a scalar operator $\phi(x)$ with dimension $\Delta_\phi$ in a Euclidean $d$-dimensional CFT takes the general form\footnote{For our notations see Appendix \ref{app:notation}.}
\be
\label{2ptCFTgen}
\langle\phi(\tau,\vec{x})\phi(0,\vec{0})\rangle_d\equiv g_d(\tau,\vec{x})=\frac{1}{r^{2\Delta_\phi}}\sum_{\{\cQ_s\}}a^{(d)}_{\cQ_s}\,{\cal G}^{(d)}_{s,\Delta_{\cQ_s}}(r,\cos\theta)\,.
\ee
The $d$-dimensional thermal conformal blocks\footnote{Perhaps a more accurate name could be restricted thermal conformal blocks, in order to distinguish them from the more complicated thermal blocks in the $S^1\times S^{d-1}$ geometry \cite{Gobeil:2018fzy}. To our knowledge the momentum space version of (\ref{2ptCFTgen}) has first appeared in \cite{Petkou:1998fb} and its coordinate version more recently in \cite{Iliesiu:2018fao,Petkou:2018ynm} and in all subsequent works on thermal two-point functions.}  are given as
\be
\label{thconfblocks}
{\cal G}^{(d)}_{s,\Delta_{\cQ_s}}(r,\cos\theta)=C_s^{\n}(\cos\theta)\left(\frac{r}{\beta}\right)^{\Delta_{\cQ_s}}
\ee
with $C_s^{\n}(t)$ the Gegenbauer polynomials and $\n=d/2-1$.
The expansion (\ref{2ptCFTgen}) arises from the OPE of $\phi$ with itself in the corresponding zero-temperature CFT, which means that it contains all the relevant conformal data of the latter theory i.e.  scaling dimensions, three-point function coefficients and information regarding the operator spectrum. The additional purely thermal information comes in the form of thermal one-point functions of the quasiprimary operators $\cQ_s$ with spin-$s$ and scaling dimension $\Delta_{\cQ_s}$ that appear in the OPE and it is contained in the constant coefficients $a^{(d)}_{\cQ_s}$.  In a field-theoretic representation the operators $\cQ_s$  are given by rank-$s$ symmetric traceless tensors and their nonzero thermal one-point functions have the form
\begin{equation}
    \langle \cQ_s\rangle_d \equiv \langle \cQ_{\mu_1\cdot\cdot\cdot\mu_s} \rangle_d = \frac{b^{(d)}_{\cQ_s}}{\beta^{\Delta^{(d)}_s}}(e_{\mu_1}e_{\mu_2}\cdot\cdot\cdot e_{\mu_s}-\text{traces})\,,\label{eq:1ptfunction}
\end{equation}
with $b^{(d)}_{\cQ_s}$ constants and $e_{\mu}$  the unit vector in $\tau$ direction. The relation of  $a^{(d)}_{\cQ_s}$ and  $b^{(d)}_{\cQ_s}$ is
\begin{equation}
    a^{(d)}_{\cQ_s} = \frac{f_{\phi\phi \cO_s}b^{(d)}_{\cQ_s}}{c_{\cQ_s}}\frac{s!}{2^s(\nu)_{s}}\,,\label{eq:atob}
\end{equation}
where $c_{\cQ_s}$, $f_{\phi\phi \cQ_s}$ are the coefficients of the two- and three-point functions $\langle \cQ_s\cQ_s\rangle$, $\langle\phi\phi\cQ_s\rangle$ respectively,  and $(a)_n=\Gamma(a+n)/\Gamma(a)$. In the thermal bootstrap approach \cite{Iliesiu:2018fao,Iliesiu:2018zlz,Barrat:2025nvu,Buric:2025anb,Buric:2025fye} one aims to find the coefficients $a^{(d)}_{\cQ_s}$ using the analytic properties of the two-point function together with the KMS condition $g(\tau+\beta,\vec{x})=g(\tau,\vec{x})$. Notice that the latter condition is not manifest in the expansion (\ref{2ptCFTgen}). Progress in the thermal conformal bootstrap has not yet been as spectacular as in its zero-temperature counterpart, but a number of encouraging results have recently emerged.

\subsection{Thermal two-point functions of massive free scalars in $d$-dimensions}

An expansion of the form (\ref{2ptCFTgen}) that satisfies the KMS condition does not necessarily represent conformal thermal two-point functions. It may as well represent thermal two-point functions of massive free scalars in dimensions $d=2L+1$, $L=0,1,2,3,..$  with Lagrangian
\be
\label{L_massive_scalar}
\cL_{d}=\frac{1}{2}\phi(-\Box+m^2)\phi\,,\,\,\,\,\,\Box=\frac{\partial^2}{\partial\tau^2}+\vec{\nabla}^2\,.
\ee
The relevant results have been presented in the series of works \cite{Petkou:2021zhg,Karydas:2023ufs,Karydas:2025tfs} and we review them here in some detail. 
Starting from  (\ref{L_massive_scalar}) the calculation is standard (see, e.g. \cite{Bellac:2011kqa}) and gives
\begin{eqnarray}
\label{phiphi_thermal_massive}
g_{d}(\tau,\vec{x}) &=& \frac{1}{\beta}\sum_{n=-\infty}^\infty\int\frac{\mathrm{d}^{d-1}\vec{p}}{(2\pi)^{d-1}}\frac{e^{-i\omega_n\tau-i\vec{p}\cdot\vec{x}}}{\omega_n^2+\vec{p}^2+m^2}=\sum_{n=-\infty }^{\infty}\int\frac{\mathrm{d}^dp}{(2\pi)^d}\frac{e^{-ip_0(\tau-n\beta)-i\vec{p}\cdot\vec{x}}}{p^2+m^2}\nonumber\\
    &=& \frac{1}{(2\pi)^{\frac{d}{2}}}\left(\frac{m}{r}\right)^{\n} K_{\n}(mr)+\frac{1}{(2\pi)^{\frac{d}{2}}}\sum_{\substack{n=-\infty \\ n\neq 0}}^{\infty}\left(\frac{m}{|X_n|}\right)^{\n} K_{\n}(m|X_n|) \,,
\end{eqnarray}
with $K_\n(z)$ the modified Bessel function of the second kind and $X^\mu_n = (\tau-n\beta,\vec{x})$. 
The finite temperature contribution $n\neq 0$ is a sum of mirror images of the corresponding zero temperature part and hence (\ref{phiphi_thermal_massive}) satisfies the KMS condition. In odd dimensions $d=2L+1$ 
we can obtain analytic results. For the zero temperature contribution we obtain
\begin{equation}
\label{zeroT}
    \frac{1}{(2\pi)^{d/2}}\left(\frac{m}{r}\right)^\nu K_\nu(mr) = \frac{1}{r^{d-2}}\left(\sum_{k=0}^\infty A_{2k}^{(d)}(m\beta) \left(\frac{r}{\beta}\right)^{2k} + \sum_{k=0}^\infty B^{(d)}_{2k+1}(m\beta) \, \left(\frac{r}{\beta}\right)^{2k+1}\right)\,,
\end{equation}
with the relevant calculations and the coefficients given in Appendix \ref{app:expansionof0T}. There is no temperature dependence here, but we have nevertheless introduced $\b$ in order to interpret the expansion (\ref{zeroT}) as contributions from spin-zero thermal conformal blocks. 

The expansion of the finite temperature part of (\ref{phiphi_thermal_massive}) in thermal conformal blocks has been achieved in \cite{Iliesiu:2018fao,Petkou:2018ynm} using the so-called inversion formula. It gives contributions from even spin $s$ and mixes only  with the second series on the r.h.s. of (\ref{zeroT}). The corresponding expansion coefficients depend on the coordinates $(\tau,\vec{x})$ and their calculation is nontrivial. Using the results of \cite{Iliesiu:2018fao,Petkou:2018ynm} the thermal two-point function takes the general form\footnote{To avoid clutter we use hereafter a reduced notation for the coefficients and the conformal blocks in thermal two-point functions.}
\be
\label{phiphi1}
g_{d}(\tau,\vec{x})=
\frac{1}{r^{d-2}}\Biggl[\sum_{k=0}^{\infty}A_{2k}^{(d)}{\cal G}^{(d)}_{0;2k}+ \sum_{\substack{s=0,2,4,..\\ t=0,1,2,..}}a^{(d)}_{\cQ_{s;t}}{\cal G}^{(d)}_{s;d-2+s+2t}\Biggl]\,.
\ee
The coefficients $a_{\cQ_{s;t}}^{(d)}$ have been explicitly calculated for $t=0$  in terms of finite sums of polylogarithms $\text{Li}_s(e^{-m\b})$ as \cite{Iliesiu:2018fao,Petkou:2018ynm}
\begin{align}
\label{aQs0}
   a^{(d)}_{\cO_{s;0}}&=\delta_{s0}\frac{(-1)^{\frac{d-1}{2}}(m\b)^{d-2}}{2(2\pi)^{\frac{d-1}{2}}(2\n)!!}+\sum_{n=0}^{d/2-3/2+s}D_n(d,s)\,(m\b)^n\,\,\text{Li}_{d-2+s-n}(e^{-m\b}) \\
   D_n(d,s)&=\frac{1}{(\n)_s(2\pi)^{\frac{d-1}{2}} 4^{s}}\frac{2^{n-1}}{n!}\frac{(d-3+2s-n)!}{\left(\frac{d-3}{2}+s-n\right)!}
\end{align}
The expansion (\ref{phiphi1}) is certainly not surprising. It is known since the seminal work of Wilson et al. \cite{Wilson:1969zs,Nachtmann:1973mr} that symmetric traceless tensors with assigned scaling dimensions constitute a useful basis for OPEs in generic QFTs. Conformal symmetry puts  order in this basis by identifying the quasiprimary operators annihilated by special conformal transformations, and fixing the structure of their descendants. However, in the thermal state the information regarding the descendants is washed out since their one-point functions vanish. In the case of the integrable massive deformations that we consider here the mass parameter $m$ does not alter the basis of quasiprimary operators, but it only affects the expansion coefficients. As a consequence, the thermal two-point function of (\ref{L_massive_scalar}) can be expanded into thermal conformal blocks that are associated with contributions of symmetric traceless tensors with well-defined scaling dimensions. Although these tensor fields do not correspond to proper operators of the massive theory, nevertheless they can be thought of as a useful conformal basis for its two-point functions. In the explicit example above this basis contains three classes of operators:
\begin{itemize}[leftmargin=10pt]
\item
Scalar operators of the form $\cQ'_{0;n}=\sigma^n$ with dimensions $\Delta_n=2n$, $n=1,2,..$. These correspond to powers of  the operator $\sigma$ which, in the conformal limit,  is the shadow operator of $\phi^2$ and has fixed scaling dimension $\Delta_\sigma=2$ for all $d$.  Their thermal one-point functions give rise to the first series on the r.h.s. of (\ref{phiphi1}). 
\item 
Leading-twist, higher-spin operators which are bilinears of the form\footnote{The twist of an operator is defined as $\frac{\Delta_{\cQ_s}-s}{2}-\Delta_\phi$. We discard total derivative terms which do not contribute to thermal one-point functions. See Appendix \ref{app:productexpansion} for the relevant formulae.} $\cQ_{s;0}\sim \phi D^{(s)}_{\mu_1\cdots\mu_s}\phi$ with assigned scaling dimensions $ \Delta^{(d)}_{\cQ_{s}} \equiv \Delta_{d,s}= d-2+s$. $D^{(s)}_{\mu_1\cdots\mu_s}$ is a symmetric and traceless tensor made out from derivatives. Their thermal one-point functions  contribute to the second series on the r.h.s. of (\ref{phiphi1}) for $t=0$. We call these operators would-be higher-spin conserved currents since their conservation is established at the conformal points where they turn into true conformal quasiprimaries. 
\item
Higher-twist, higher-spin operators of the form $\cQ_{s;t}\sim \phi D^{(s)}_{\mu_1\cdots\mu_s}\Box^t\phi$ with assigned  scaling dimensions $ \Delta^{(d)}_{\cQ_{s;t}} = \Delta_{d,s}+2t$. They give rise to the coefficients in the second series in (\ref{phiphi1}) for $t\neq 0$. 
 \end{itemize}

\subsubsection{The higher-twist dressing of the expansion coefficients}
The  construction of the higher-twist operators $ \cQ_{s;t}$ requires using the equation of motion $\Box\phi=m^2\phi$ multiple times. This implies that the corresponding coefficients have the form
\begin{equation}
    a_{\cQ_{s;t}}^{(d)} = C(d,t,s)(\beta m)^{2t}a_{\cQ_{s;0}}^{(d)}\,.\label{highertwistansatz}
\end{equation}
From \eqref{phiphi_thermal_massive}
 and a property of the Bessel functions we obtain\footnote{Note that this relation also holds for even dimensions $d>4$.}
\begin{equation}
\label{ggg}
    \frac{\partial}{\partial m^2}g_{d}(\tau,\vec{x}) = -\frac{1}{4\pi}g_{d-2,1}(\tau,\vec{x})\,.
    \end{equation}
Expanding both sides of  (\ref{ggg})  using \eqref{2ptCFTgen} we find
\begin{equation}
    \frac{1}{\beta^2}\sum_{\cQ_{s;t}} \left(\frac{\partial}{\partial m^2}a_{\cQ_{s;t}}^{(d)}\right)\left(\frac{r}{\beta}\right)^{\Delta_{\cQ_{s;t}}-d+2}C^{\n}_{s}\left(\cos\theta\right) = -\frac{1}{4\pi}\sum_{\cQ'_{s;t}} a_{\cQ'_{s;t}}^{(d-2)}\left(\frac{r}{\beta}\right)^{\Delta_{\cQ'_{s;t}}-d+4}C^{\n-1}_{s}\left(\cos\theta\right)\,.
\end{equation}
Using the relation of the Gegenbauer polynomials\footnote{Notice  that $C_0^{\lambda}(\cos\theta)=1$ and $C_n^{\lambda}(\cos\theta)=0$ for $n<0$. }
\begin{equation}
\label{gegen_uplift}
    C^{\n}_{s}(\cos\theta) = \frac{\n}{\n+s}\left[C^{\n+1}_s(\cos\theta) - C^{\n+1}_{s-2}(\cos\theta)\right]\,,
\end{equation}
and the recursive formula of \cite{Karydas:2023ufs,Karydas:2025tfs} for the coefficients (\ref{aQs0}) 
\begin{equation}
\label{recursion}
    a_{\cQ_{s+2;0}}^{(d)} = \frac{2\pi}{d-2}a_{\cQ_{s;0}}^{(d+2)} + \frac{(\beta m)^2}{(d-2+2s)(d+2s)}a_{\cQ_{s;0}}^{(d)}\,,
\end{equation}
we find for $t=0$
\begin{equation}
    -\frac{1}{\beta^2}\frac{\partial}{\partial m^2}a_{\cQ_{s;0}}^{(d)} = \frac{1}{4\pi}\frac{d-4}{d-4+2s}a_{\cQ_{s;0}}^{(d-2)}\,.
\end{equation}
and for $t>0$
\begin{equation}
    -\frac{1}{\beta^2}\frac{\partial}{\partial m^2}a_{\cQ_{s;t}}^{(d)} = \frac{1}{4\pi}\left(\frac{d-4}{d-4+2s}a_{\cQ_{s;t}}^{(d-2)} - \frac{d-4}{d+2s}a_{\cQ_{s+2;t-1}}^{(d-2)}\right)\,.
\end{equation}
Using then (\ref{recursion}) we obtain the following recursive relations
\begin{eqnarray}
    tC(d,t,s) &=& \frac{1}{2(d+2s)}C(d-2,t-1,s+2)\,,\\
    C(d,t,s)&=& C(d-2,t,s) - \frac{C(d-2,t-1,s+2)}{(d+2s)(d-2+2s)}\,.
\end{eqnarray}
Notice that $C(d,t=0,s)=1$. We can then solve the recursive relations above to obtain 
\begin{equation}
\label{Cresult}
    C(d,t,s) = \frac{1}{4^tt!\left(\frac{d}{2}+s\right)_t}\,.
\end{equation}
Finally, using the following identity
\begin{equation}
\label{twistsum}
\sum_{t=0}^{\infty}
\frac{(mr)^{2t}}
{4^t\,t!\left(\frac{d}{2}+s\right)_t} 
= {}_0 F_1 \!\left( \frac{d}{2} + s \,;\, \frac{(mr)^2}{4} \right)
= \Gamma\!\left(\frac{d}{2}+s\right)
\left(\frac{2}{mr}\right)^{\frac{d}{2}+s-1}
I_{\frac{d}{2}+s-1}(mr)
\end{equation}
where $_0F_1(a;z)$ is a confluent hypergeometric function and $I_\a(z)$ is the modified Bessel function of the first kind, we can perform the sum over the twists and bring (\ref{phiphi1})  into the form
\be
\label{phiphi11}
g_{d,1}(r,\vec{x})=
\frac{1}{r^{d-2}}\left[\sum_{k=0}^{\infty}A_{2k}^{(d,1)}{\cal G}^{(d)}_{0;2k}+ \sum_{s=0,2,4,..}{\cal B}^{(d,1)}_{s}\,{\cal G}^{(d)}_{s;\Delta_{d,s}}\right]\,,
\ee
with 
\be
\label{adressed}
{\cal B}^{(d,1)}_{s}(m\b,mr)={}_0 F_1 \!\left( \frac{d}{2} + s \,;\, \frac{(mr)^2}{4} \right)a^{(d)}_{\cQ_{s;0}}.
\ee
In other words, the thermal two-point function (\ref{phiphi1}) can be expanded in conformal blocks of  would-be higher-spin currents having dimensions $\Delta_{d,s}=d-2+s$, with dressed coefficients that encode the higher-twist contributions for each spin. We call the would-be higher-spin operators the pillars on which the higher-twist columns are built for each spin. The pillars correspond to the operators that survive in the massless limit. 

In Appendix \ref{app:d=1}  we consider the spatial coincident limit of the thermal two-point functions and also the correspondence with the harmonic oscillator for $d=1$. In Appendix \ref{app:productexpansion} we present some details for the structure of the higher-spin tensors and an alternative algebraic method to arrive at the results (\ref{Cresult}) and (\ref{adressed}).
\section{From massive free scalars to thermal CFTs}\label{sec:gapeq}

\subsection{The thermal gap equation}

As we have mentioned above, although  (\ref{phiphi1}) admits an expansion in thermal conformal blocks and also satisfies the KMS condition, this is not enough to associate it to the thermal CFT since the mass parameter breaks scale invariance explicitly. This could be rectified if $m$ was somehow tied to $\beta$ i.e. by fixing the dimensionless parameter $\lambda=\beta m$. However, if $\lambda$ were allowed to be arbitrary it would imply the existence of a fixed line of thermal CFTs continuously connected under a massive deformation to the free massless thermal theory. Such a line would be projected to a corresponding fixed line among zero temperature CFTs and clearly this is not the case here. Nevertheless, since we have implicitly assumed the presence of an energy-momentum tensor in the expansion (\ref{phiphi11}), namely a spin-2 operator with scaling dimension $d$, there exists a general condition that fixes the parameter $\lambda=\beta m$ such that the thermal two-point function becomes conformal. 

Consider the coefficient $a^{(d)}_{\cQ_{2;0}}$ in (\ref{phiphi1}). This is the contribution of a symmetric traceless spin-2 operator $\cQ_{\m\n}\equiv \cQ_2$, with scaling dimension $\Delta_{\cQ_2}=d$, which is not the energy-momentum tensor of the massive theory (\ref{L_massive_scalar}). From (\ref{eq:1ptfunction}) and (\ref{eq:atob}) $a^{(d)}_{\cQ_{2;0}}$ depends on the ratio $f_{\phi\phi\cQ_2}/c_{\cQ_2}$ and on the thermal two-point function coefficient $b_{\cQ_2}=d\langle\cQ_{\tau\tau}\rangle/(d-1)$. Our assumption is that the massive deformation affects only the latter and that we use the general free field theory result\footnote{We use the sign convention of \cite{Petkou:1994ad} where $f_{\phi\phi T}<0$.} $f_{\phi\phi\cQ_s}/c_{\cQ_s}=-1/s!$  (see \cite{Skvortsov:2015pea,Benedetti:2023pbt} and the Appendix \ref{app:productexpansion}) for the would-be higher-spin currents to obtain 
\be
\label{cQtt}
a^{(d)}_{\cQ_{2;0}}=-\frac{\b^d \langle \cQ_{\tau\tau}\rangle}{(d-1)(d-2)}\,.
\ee
This is  related to the partition function $\cZ_d$ and free energy  $\cF_d$ of (\ref{L_massive_scalar}) as follows. Dimensional analysis requires that\footnote{Here and in the following $\cF_d$   is the renormalized result which is obtained after the subtraction of the zero temperature UV divergences. With our definition $\cC_d(m\b)<0$ for systems with positive pressure.} 
\be
\label{Fd}
-\ln\cZ_d=\beta\cF_d(\beta)=\frac{V_{d-1}}{\b^{d-1}}\cC_d(m\b)\,,
\ee
where $V_{d-1}$ is the spatial volume and $\cC_d(m\b)$ is a dimensionless function.  Crucially, the  results of \cite{Karydas:2025tfs} associate a scalar thermal coefficient in $d+2$-dimensions with the free energy of (\ref{L_massive_scalar}) as\footnote{This follows from the first of Eq. (5.7) in \cite{Karydas:2025tfs}, if we correct the obvious misprint $\a^{2L}\rightarrow \a^{2(L-1)}$ and take into account that the formulae in that work correspond to two scalar fields.}
\be
\label{atoC}
a^{(d+2)}_{\cQ_{0;0}}(m\b)= -\frac{1}{2\pi}\cC_d(m\b)\,.
\ee
Using  (\ref{recursion}) we find
\be
\label{QttCd}
-\beta^d\langle\cQ_{\tau\tau}\rangle=(1-d)\cC_d+\frac{d-1}{d}(\b m)^2a^{(d)}_{\cQ_{0;0}}\,.
\ee

On the other hand  the standard definitions of the energy density $\cal{E}$ and pressure $\cal{P}$ in terms of the thermal averages\footnote{The normalized thermal average of an operator $\hat{\cO}$ is defined as 
$\langle\hat{\cO}\rangle =\frac{1}{\cZ}\Tr\left[\hat{\cO}e^{-\b\hat{H}}\right]$. } of the Euclidean energy-momentum tensor are
\be
\label{HTF}
\langle\hat{H}\rangle =V_{d-1}{\cal E}=-V_{d-1}\langle T_{\tau\tau}\rangle\,,\,\,\,\,V_{d-1}\langle T_{ii}\rangle =V_{d-1}{\cal P}=-\cF_d\,.
\ee
Deriving (\ref{Fd}) we find
\be
\label{Ttrace}
-\b^d\langle T_{\tau\tau}\rangle =(1-d)\cC_d+\beta\frac{\partial}{\partial\b}\cC_d(m\b)
\ee
Similar results have been derived in \cite{CastroNeto:1992ie, LeClair:2005aa}. Trading the $\beta$ and $m$-derivatives we have 
\be
\label{avrgphi2}
\frac{\partial}{\partial m^2}\ln\cZ_d=-\frac{1}{2}\b V_{d-1}\langle\phi^2\rangle_d\,\Rightarrow\, \b\frac{\partial}{\partial\b}\cC_d(m\b)=(\b m)^2\b^{d-2}\langle\phi^2\rangle_d=(\b m)^2a^{(d)}_{\cQ_{0;0}}\,,
\ee
which leads to 
\be
\label{conformal_condition3}
-\b^d\langle T_{\tau\tau}\rangle =(1-d)\cC_d+(\b m)^2a^{(d)}_{\cQ_{0;0}}
\,.
\ee
Next, if we identify the traceless part of the energy-momentum tensor $T_{\m\n}$  with $\cQ_{\m\n}$ as
\be
\label{TQ}
\cQ_{\m\n}=T_{\m\n}-\frac{1}{d}\delta_{\m\n}T^{\rho}_{\,\,\,\rho}\,,
\ee
we will obtain from (\ref{QttCd}) and  (\ref{conformal_condition3}) the following relation
\be
\label{traceT}
\b^d\langle T^{\rho}_{\,\,\,\rho}\rangle=-(\b m)^2a^{(d)}_{\cQ_{0;0}}\,.
\ee
Therefore, even if the energy-momentum tensor is not traceless at zero temperature, the thermal average of its trace can be zero if  there exists a value of $m_*$ such that it satisfies the gap equation
\be
\label{gap_eq1}
m_*^2a^{(d)}_{\cQ_{0;0}}=\b^{d-2}m_*^{2}\langle\phi^2\rangle_d=0\,.
\ee
In our discussion we have taken the temperature to be fixed (i.e. it could be set to 1), and considered $m$ as the adjustable parameter of the theory since the latter has a clear zero temperature interpretation. On the other hand one might take $m$ fixed and vary $\b$, in which case (\ref{gap_eq1}) would resemble the well-known effect of high-temperature restoration of a global symmetry (e.g. see \cite{Dolan:1973qd}) -  dilatation  in this case. However the mass deformation explicitly breaks conformal invariance so the trace of the energy-momentum tensor is not an order parameter. Nevertheless,  for fixed $\b$ and different values of $\lambda=m_*\b$ that satisfy the gap equation (\ref{gap_eq1}), the thermal two-point function (\ref{phiphi1}) is associated with different thermal CFTs. In that  sense the thermal trace of the energy-momentum tensor can be used as a diagnostic for the conformal invariance of the corresponding zero-temperature theories.  

In the case of (\ref{gap_eq1}) we have two classes of solutions. Either $m_*=0$, in which case the thermal average $\langle\phi^2\rangle_d$ can be nonzero, or $m_*\neq 0$, in which case the thermal fluctuations should cancel the quantum vacuum fluctuations such that $\langle\phi^2\rangle_d$ vanishes.  The nonzero value of $m_*$ that satisfies the gap equation  is sometimes called the thermal mass. A minor novelty here is that we have arrived at the  gap equation using the expansion of the two-point function in thermal conformal blocks, and crucially (\ref{atoC}) which relates the expansion coefficients to the free energy. A nontrivial solution for (\ref{gap_eq1}) indicates that there exist nontrivial CFTs at zero temperature that dictate the physics at the high-temperature critical point. 

\subsection{The $m_*=0$  solution of the gap equation}

For $m_*=0$ the thermal two-point function (\ref{phiphi1}) takes the form 
\be
\label{phiphi0}
g_{d}(\tau,\vec{x}) =\sum_{n=-\infty}^\infty\frac{C_\phi(d,1)}{[(\tau-n\beta)^2+\vec{x}^2)]^{\Delta_\phi}}=\frac{C_\phi(d,1)}{r^{d-2}}\left[1+\sum_{s=0,2,4,..}^{\infty}2\zeta(d-2+s){\cal G}_{s,\Delta_{d,s}}^{(d)}\right]\,.
\ee
This is associated with the zero-temperature CFT$_d$ of massless free scalars $\phi(x)$ with scaling dimensions $\Delta_\phi=d/2-1$ satisfying the periodic boundary condition $\phi(\tau+\beta,\vec{x})=\phi(\tau,\vec{x})$. The expansion (\ref{phiphi0}) receives contributions just from higher-spin conserved currents  $\cQ_{s;0}\sim \phi D^{(d)}_{\mu_1\cdots\mu_s}\phi$, with even spin and scaling dimensions $\Delta_{d,s}$. Higher-twist, higher-spin operators are absent due to the e.o.m. $\Box\phi=0$. This matches the operator content of a massless free zero-temperature CFT$_d$.

The thermal conformal block coefficients are $a^{(d)}_{\cQ_{s;0}}=2C_{\phi}(d,1)\zeta(d-2+s)$.  
 Using Eq. (3.23) of \cite{Karydas:2025tfs} we find $\cC_d{(0)}=-2\zeta(d)/S_d=-2\pi a^{(d+2)}_{\cQ_{0;0}}$  and hence (\ref{atoC}) is verified. Notice that here  $a^{(d)}_{\cQ_{0;0}}\sim \langle\phi^2\rangle_d\neq 0$. The case $d=3$ is special since $\langle\phi^2\rangle_3\sim\zeta(1)$ diverges. This can be traced to the logarithmic divergence of the zero mode in the coincident limit of (\ref{phiphi_thermal_massive})  and corresponds to an IR divergence. It is also connected to the Mermin-Wagner-Coleman theorem for the absence of spontaneous symmetry breaking of continuous symmetries in two dimensions \cite{Rosenstein:1989sg}.

\subsection{The $m_*\neq 0$ solutions of the gap equation}

For $m_*\neq 0$ the gap equation (\ref{gap_eq1}) takes the form  
\begin{equation}
\label{gap_eq2}
   \sum_{n=0}^{\frac{d-3}{2}}\frac{(d-3-n)!(2m_*\beta)^n}{(\frac{d-3}{2}-n)!n!}\mathrm{Li}_{d-2-n}(e^{-\beta m_*}) + \frac{(-1)^{\frac{d-1}{2}}2^{\frac{d-5}{2}}(m_*\beta)^{d-2}}{(d-2)!!} = 0\,.
\end{equation}
The polylogarithms $\mathrm{Li}_s(z)$ take positive values when $0<e^{-\b m}<1$ and (\ref{gap_eq2}) has real solutions only if the second term is negative, i.e. for $d=4n+3$, $n=0,1,2,3,..$. When $d=4n+5$ all solutions of (\ref{gap_eq2}) are complex. The careful reader will certainly notice that (\ref{gap_eq2}) is just an elaborate way to present a well-known result - namely, the second term on the lhs of (\ref{gap_eq2}) is just the zero temperature quantum vacuum fluctuations of the scalar field $\langle\phi^2\rangle_d$ in odd dimensions and alternates in sign. The first term on the lhs of (\ref{gap_eq2}) represents the thermal fluctuations. When these two contributions cancel each other we have a crossover between quantum and classical regimes. At this point (\ref{phiphi1}) can be associated with a zero-temperature CFT$_d$ and becomes a conformal thermal two-point function. The operator spectrum of the CFT$_d$ is as follows: it includes higher-spin conserved currents of the form $\cQ_{s;0}$,  it includes higher-twist, higher-spin operators of the form $\cQ_{s;t}$ for $t>1$, it includes powers of the shadow field $\sigma$ but it does not include $\phi^2$ and all other higher-twist scalar operators of the form $\phi\Box^t\phi$  who have dropped out from the spectrum.

Our idea is that the massive deformations above explore the space of thermal QFTs near its massless free point and they find nearby thermal CFTs when the gap equation (\ref{gap_eq1}) is satisfied.  This picture should have a projection in the space of CFTs at zero temperature, but we cannot claim that it corresponds to an RG flow.  This is indeed the case in $d=3$ where the thermal two-point function (\ref{phiphi1}) for $m_*\neq 0$ is associated with the nontrivial fixed point of the $O(N)$ vector model for $N\mapsto\infty$ and the corresponding operator spectra match. The latter theory is obtained from the massless free case by a Hubbard-Stratonovich transformation for large-$N$. 

We would expect that for each nontrivial solution of the gap equation (\ref{gap_eq1}) one should be able to associate a corresponding CFT$_d$ at zero temperature. This raises the question of the physical interpretation of the complex solutions of (\ref{gap_eq1}). It has been claimed in \cite{Petkou:2018ynm} that such solutions indicate an instability in the putative nontrivial large-$N$ CFTs in $d=5,9,..$ dimensions. The argument goes as follows. One way to study nontrivial large-$N$ CFTs in odd dimensions is to view them as IR fixed points of UV complete massless free CFTs via the $\epsilon$-expansion. Consider the classically marginal interaction of the massless free Lagrangian in $d$-dimensions
\be
\label{L1}
\cL=\frac{1}{2}\phi(-\Box)\phi +g\phi^{\frac{2d}{d-2}}\,.
\ee
The idea is then to consider (\ref{L1}) in  $d'=d-\epsilon$ dimensions with $\epsilon>0$ where the interaction becomes relevant and study possible critical points there, taking the limit $\epsilon\rightarrow 1$ at the end. This procedure can be combined with some other parameter such as large-$N$ and at high orders gives reliable results. Here we use it only as a proof of concept. Introducing the Hubbard-Stratonovich field $\sigma$ this can be written as
\be
\label{LHS}
\cL_{HS}=\frac{1}{2}\phi(-\Box+\sigma)\phi -\lambda\sigma^{\frac{d}{2}}\,,\,\,\,\,\,\lambda =\frac{1}{d}\left(\frac{d-2}{2d}\frac{1}{g}\right)^{\frac{d-2}{2}}\,.
\ee
The latter is quadratic in $\phi$ and yields an effective action for $\sigma$ of the form
\be
\label{Veff}
V_{eff}=-\frac{1}{2}\Tr\ln(-\Box+\sigma)-\lambda\sigma^{\frac{d}{2}} \sim -\sigma^{\frac{d}{2}}(\ln|\sigma| -\lambda)\,.
\ee
This is an even function of $\sigma$ for $d=4,8,12,..$ while it is an odd function of $\sigma$ for $d=6,10,14,..$. In the first case we expect global minima, while in the second we do not. This is an indication e.g. that in $d=5$ the large-$N$ nontrivial fixed point does have some non-perturbative instability, something that has been studied extensively in \cite{Fei:2014yja,Giombi:2019upv}. For a similar discussion on fermionic theories see also \cite{David:2023uya}.

\section{Thermal dimensional uplifts and higher-derivative CFTs}\label{sec:uplift}

\begin{figure}[h!]
    \centering    \includegraphics[width=1\linewidth]{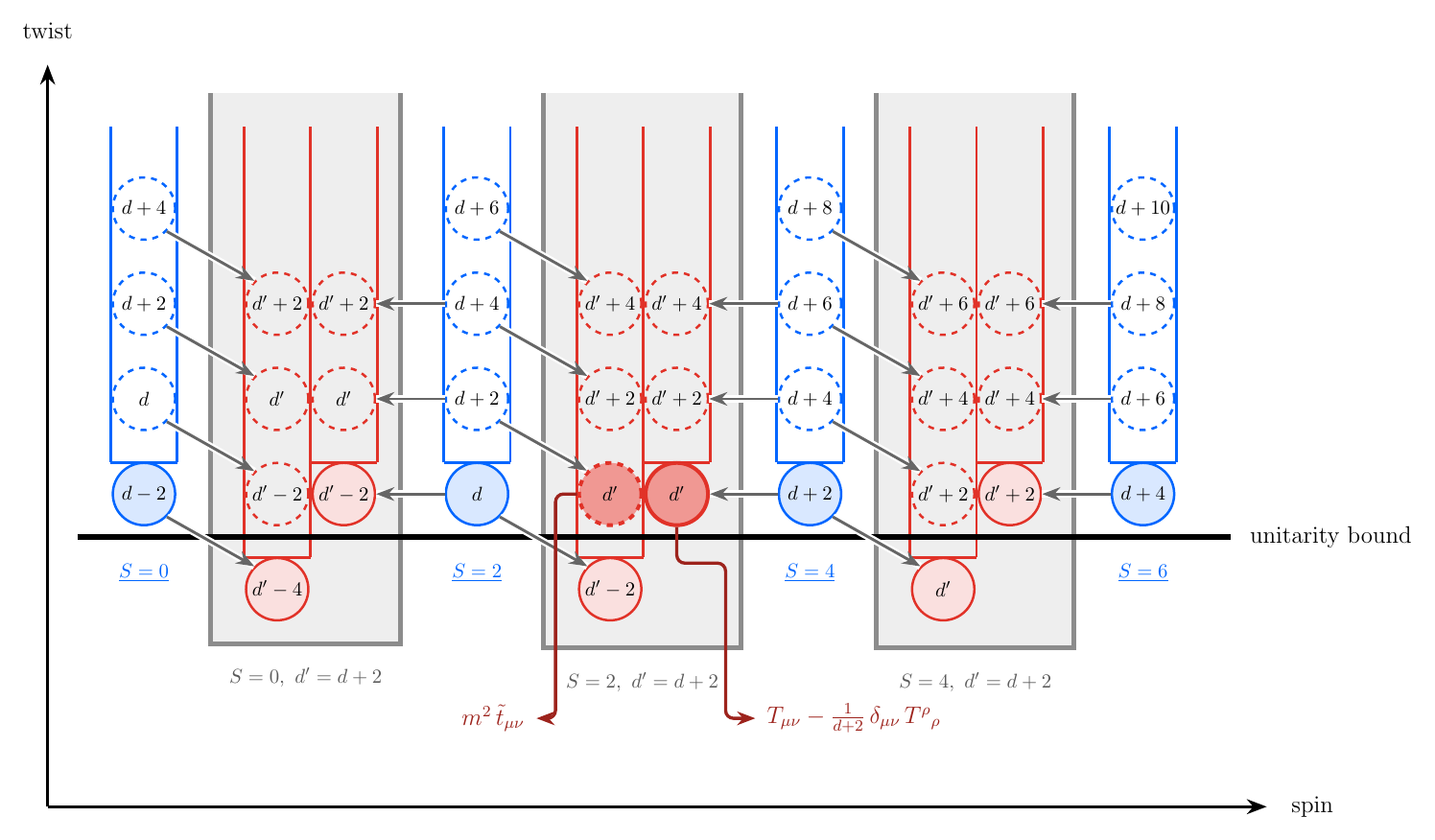}
    \caption{The thermal dimensional uplift: Blue columns correspond to the $d$-dimensional theory and the red columns to the $d'$-dimensional theory with $d'=d+2$. Blobs represent operators of given spin-$S$ (horizontal axis) and
scaling dimension (shown inside); height measures the twist above the unitarity
bound of the respective theory. Solid blobs mark the pillar operators at the
base of each column and dashed blobs the higher-twist operators built on them.
Arrows connect blobs of equal scaling dimension and implement (4.2): blue
spin-$S$ to red spin-$S$ (coefficient $\mathcal C_S$), and blue spin-$(S+2)$ to
red spin-$S$ (coefficient $\mathcal B_S$). The darker blobs in the $S=2$ box are
$m^2\tilde t_{\mu\nu}$ and $T_{\mu\nu}-\frac{1}{d+2}\delta_{\mu\nu}T^{\rho}{}_{\rho}$,
both of spin-2 and dimension $d+2$; their sum is $\tilde Q_{\mu\nu}$ of (\ref{QTt}).}
    \label{fig: conf integrals}
\end{figure}

The compact expansion (\ref{phiphi11}) of the massive thermal two-point function motivates us to apply a straightforward dimensional uplift. By virtue of (\ref{gegen_uplift}) the  thermal blocks of the would-be higher spin currents in $d$-dimensions can be expanded in $d+2$-dimensional thermal blocks. 
This allows us to interpret $g_{d,1}(r,\vec{x})$ as a thermal two-point function of a $d+2$-dimensional higher-derivative massive theory since it can be written as 
\be
\label{phiphi111}
g_{d,1}(r,\vec{x})=
\frac{1}{r^{d-2}}\left[\sum_{k=0}^{\infty}{\cal A}_{2k}^{(d+2,2)}{\cal G}^{(d+2)}_{0;2k}+ \sum_{s=0,2,4,..}\left\{ {\cal C}^{(d+2,2)}_s {\cal G}^{(d+2)}_{s;\Delta_{d+2,s}-2}+ {\cal B}^{(d+2,2)}_s{\cal G}^{(d+2)}_{s,\Delta_{d+2,s}}\right\}\right]\,,
\ee
where
\be
\label{D_coeff}
{\cal A}^{(d+2,2)}_{2k}=A_{2k}^{(d,1)}\,,\,\,\,\cC^{(d+2,2)}_{s}=\frac{d-2}{d-2+2s}{\cal B}^{(d,1)}_{s}\,,\,\,\,{\cal B}^{(d+2,2)}_s=-\frac{d-2}{d+2+2s}{\cal B}^{(d,1)}_{s+2}\,.
\ee
Now (\ref{phiphi111}) represents an alternative expansion of the same thermal two-point function in which we have two classes of pillar operators: on the one hand the would-be conserved  currents in $d+2$-dimensions with  $\Delta_{d+2,s}=d+s$, and on the other hand symmetric traceless tensors with scaling dimensions $\Delta_{d+2,s}-2$ that lie below the unitarity bound in $d+2$-dimensions. These are the operators that survive in the massless limit. The coefficients ${\cal C}_s^{(d+2,2)}$ and ${\cal B}_s^{(d+2,2)}$ encode the infinite columns of higher-twist operators which are built on top of each pillar operator. This uplift introduces a mixing  between operators with the same spin but different twists, as we will see in the explicit example of spin-2 below. Fig. \ref{fig: conf integrals} is a sketch of the thermal dimensional uplift procedure described above. 

The spectrum of operators that appears in (\ref{phiphi111}) matches that of a higher derivative theory in $d+2$-dimensions. Indeed, consider the class of higher-derivative Lagrangians\footnote{A nonexclusive list of recent works in higher-derivative CFTs is \cite{Safari:2017tgs,Safari:2017irw,Safari:2021ocb,Stergiou:2022qqj,Guo:2023qtt,Herzog:2024zxm,Guo:2025edk,Guo:2026vmq}.}
\be
\label{L_massive_scalar_1}
\cL_{d,k}=\frac{1}{2}\phi(-\Box+m^2)^k\phi\,, k=1,2,3,...
\ee 
Following (\ref{phiphi_thermal_massive}) their thermal two-point functions are given by
\begin{eqnarray}
\label{phiphi_thermal_massive_k}
g_{d,k}(\tau,\vec{x}) 
 = \frac{1}{2^{k-1}\Gamma(k)(2\pi)^{\frac{d}{2}}}\sum_{n=-\infty }^{\infty}\left(\frac{m}{|X_n|}\right)^{\frac{d}{2}-k} K_{\frac{d}{2}-k}(m|X_n|) \,,
\end{eqnarray}
and using known properties of the Bessel function we obtain for $k=2$ 
 \begin{equation}
\label{ggg1}
   g_{d,1}(\tau,\vec{x})=4\pi \,g_{d+2,2}(\tau,\vec{x})\,.
\end{equation}
We can now use the uplifted two-point function  (\ref{phiphi111}) in order to elucidate part of the structure of the higher-derivative theories with Lagrangians (\ref{L_massive_scalar_1}) for $k=2$.

\subsection{The dimensional uplift of massless thermal two-point functions}
Using (\ref{gegen_uplift}) we can expand (\ref{phiphi0}) in thermal conformal blocks in $d+2$ dimensions as
\be
\label{phiphi01}
g_{d+2,2}(\tau,\vec{x})=\frac{1}{4\pi}g_{d,1}(\tau,\vec{x})=\frac{1}{r^{d-2}}\left[C_\phi(d+2,2)+\sum_{\substack{s=0,2,4,..\\ t=0,1}}^{\infty}\!\!\!\tilde{a}^{(d+2)}_{\cQ_{s;t}}\,{\cal G}^{(d+2)}_{s,\Delta_{d+2,s}-2+2t}\right]\,,
\ee
where
\be
\label{ad01}
C_\phi(d+2,2)=\frac{1}{4\pi}C_\phi(d,1)\,,\,\,\,\tilde{a}^{(d+2)}_{\cQ_{s;t}}=(-1)^t\frac{\n}{\n+s+2t}2C_\phi(d+2,2)\zeta(d-2+s+2t)\,.
\ee
Formula (\ref{phiphi01}) corresponds to the conformal thermal two-point function of the scalar operator $\phi$ in $d+2$-dimensions arising from the higher-derivative Lagrangian
\be
\label{L01}
\cL_{d+2k-2,k}=\frac{1}{2}\phi(-\Box)^k\phi\,,
\ee
for $k=2$. Such theories have been considered for general integer $k\geq 2$ in many places and a comprehensive work is \cite{Brust:2016gjy}. Our results show that the thermal two-point functions of the following sequence of higher derivatives massless theories 
\be
\label{LLL}
\cL_{d,1} \rightarrow \cL_{d+2,2}\rightarrow \cL_{d+4,3}\rightarrow \cdots\rightarrow\cL_{d+2k-2,k}\rightarrow\cdots\,,
\ee
are equivalent. 
One can check that (\ref{ad01}) coincides with Eq. (4.7) of \cite{Iliesiu:2018fao}. 

The scaling dimension of $\phi$ remains $\Delta_\phi=d/2-1$ throughout the uplift, which is however below the unitarity bound for $k>1$. These $\Box^k$ theories contain $k$ Regge, or equivalently higher-twist, trajectories of higher-spin operators with dimensions $\Delta_{\cQ_{s;n-1}}=d-2+s+2(n-1)$ for $n=1,..k$.  Their dimensions are below the unitarity bound, except for the highest-twist trajectory with $n=k$ which contains the usual higher-spin currents including the energy-momentum tensor. The latter corresponds to the operator $\cQ_{2,k-1}\equiv T_{k-1}$ which using \eqref{phiphi_thermal_massive_k} and repeatedly \eqref{gegen_uplift} gives rise to the coefficient 
\begin{equation}
\label{aT}
    \a^{(d+2k-2)}_{\cQ_{2,k-1}} = (-)^{k-1}2C_\phi(d+2k-2,k)\zeta(d+2k-2)\frac{\Gamma(\frac{d}{2}+k+1)\Gamma(\frac{d}{2}+k-2)}{\Gamma(\frac{d}{2}+2k)\Gamma(\frac{d}{2}-1)}\,,
\end{equation}
where
\begin{equation}
    C_\phi(d+2k-2,k) = \frac{1}{(4\pi)^{k-1}\Gamma(k)}C(d,1)\,.
\end{equation}
Notice that this coefficient comes from the spin-$s+2(k-1)$  contribution in the original $d$-dimensional two-point function. On the other hand from (\ref{eq:atob})  the coefficient of the energy-momentum tensor in $d+2k-2$-dimensions satisfies 
\begin{equation}
    \a^{(d+2k-2)}_{\cQ_{2,k-1}}= \frac{f^{(d+2k-2)}_{\phi\phi T_{k-1}}b^{(d+2k-2)}_{T_{k-1}}}{c^{(d+2k-2)}_{T_{k-1}}}\frac{2}{(d+2k-4)(d+2k-2)}\,.\label{eq:attobt}
\end{equation}
Since $T_{k-1}$ is a conserved current its three-point function coefficient $f_{\phi\phi T_{k-1}}$ can be computed using a Ward identity \cite{Petkou:1994ad,Skvortsov:2015pea,Benedetti:2023pbt} which gives
\begin{equation}
    f_{\phi\phi T_{k-1}} = -C_\phi(d+2k-2,k)\frac{d+2k-2}{d+2k-3}\frac{\Delta_\phi}{S_{d+2k-2}}\,,\label{eq:3pt}
\end{equation}
The one-point function coefficient $b^{(d+2k-2)}_{T_{k-1}}$ is proportional to the free energy of the theory (\ref{L01}), which is $k$-times the free energy of the standard  $k=1$ massless CFT in $d+2k-2$ dimensions (see e.g. \cite{Filothodoros:2018pdj}) and we obtain 
\begin{equation}
    b^{(d+2k-2)}_{T_{k-1}} = -\frac{2k(d+2k-2)}{S_{d+2k-2}}\zeta(d+2k-2)\,.\label{eq:bt}
\end{equation}
Combining \eqref{eq:attobt}, \eqref{eq:3pt} and \eqref{eq:bt} we  find that the energy-momentum two-point function coefficient is given by
\begin{equation}
     c^{(d+2k-2)}_{T_{k-1}} = (-)^{k-1}\frac{(d+2k-2)k\Gamma(\frac{d}{2}+2k)\Gamma(\frac{d}{2})}{(d+2k-3)\Gamma(\frac{d}{2}+k+1)\Gamma(\frac{d}{2}+k-1)S_{d+2k-2}^2}\,.\label{eq:centralcharge}
\end{equation}
This coincides with the result of \cite{Osborn:2016bev,Guerrieri:2016whh} and it is obtained here rather economically using  properties of  massless thermal two-point functions. Nevertheless, we find it interesting that information such as the value of $ c^{(d+2k-2)}_{T_{k-1}}$ for higher-derivative theories (\ref{L01}) is  included in the higher-spin operators of the unitary $k=1$ theory  and can be obtained by the procedure of dimensional uplift. 

\subsection{The dimensional uplift of massive thermal two-point functions}

We ask next what is the condition that determines the conformal invariance of the higher-derivative massive thermal two-point function. We focus on the case $k=2$ and seek to understand how the energy-momentum tensor enters the expansion (\ref{phiphi111}). For that we will consider the conformal block ${\cal G}_{2;d+2}^{(d+2)}$ which is the contribution of a spin-2, dimension $d+2$ symmetric traceless tensor $\tilde{Q}_{\m\n}\equiv \tilde{Q}_2$. The coefficient of this term, call it $\alpha^{(d+2)}_{\tilde{Q}_{2}}$,  receives contributions  both from ${\cal C}_s^{(d+2,2)}$ and ${\cal B}_2^{(d+2,2)}$ in (\ref{phiphi111}). Using (\ref{adressed}) and (\ref{D_coeff}) we obtain
\be
\label{a_coeff1}
\alpha^{(d+2)}_{\tilde{Q}_{2}}=\frac{1}{4\pi}\left(\frac{d-2}{d+2}a^{(d)}_{\cQ_{2;1}}-\frac{d-2}{d+6}a^{(d)}_{\cQ_{4;0}}\right)\,.
\ee
This is a linear combination of coefficients of the original unitary $d$-dimensional theory, and in particular it contains a spin-4 contribution of the latter. Using (\ref{recursion}) we can express it just in terms of leading-twist,  spin-zero coefficients as
\be
\label{a_coeff2}
\alpha^{(d+2)}_{\tilde{Q}_{2}}=-\frac{2\pi}{2d(d+6)}a^{(d+4)}_{\cQ_{0;0}}+\frac{(d-2)(m\beta)^2}{4d(d+2)(d+6)}a^{(d+2)}_{\cQ_{0;0}}+\frac{(m\b)^4}{8\pi d(d+2)(d+6)}a^{(d)}_{\cQ_{0;0}}\,.
\ee
The general form of this coefficient can be found from (\ref{eq:atob}) and (\ref{cQtt}) to be
\be
\label{a_coeff3}
\alpha^{(d+2)}_{\tilde{Q}_{2}}=\frac{f_{\phi\phi\tilde{Q}_2}}{c_{\tilde{Q}_2}}\frac{2\beta^{d+2}\langle\tilde{Q}_{\tau\tau}\rangle}{d(d+1)}=\frac{\beta^{d+2}\langle\tilde{Q}_{\tau\tau}\rangle}{4d(d+1)(d+6)}\,.
\ee
To derive the third term in (\ref{a_coeff3}) we used the massless theory results (\ref{eq:3pt}) and (\ref{eq:centralcharge}) for $k=2$. This is consistent with our assumption that the mass deformation changes only the thermal one-point functions and not the other parameters of the underlying CFT. Notice the sign difference with (\ref{cQtt}). 
Using  relations such as (\ref{Fd}) and (\ref{atoC}) in $d+2$-dimensions we can bring (\ref{a_coeff2}) to the form
\be
\label{a_coeff4}
\beta^{d+2}\langle\tilde{Q}_{\tau\tau}\rangle = 2(d+1)C_{d+2}(m\b)+ \frac{(d+1)(d-2)}{d+2}(m\b)^2a^{(d+2)}_{\cQ_{0;0}}+\frac{1}{2\pi}\frac{d+1}{d+2}(m\b)^4a^{(d)}_{\cQ_{0;0}}\,.
\ee
On the other hand, the energy-momentum tensor of a theory with Lagrangian (\ref{L_massive_scalar_1})  satisfies the general formula  (\ref{conformal_condition3}) when we take into account that the corresponding free energy must be multiplied by $k$. In our case taking into account  (\ref{avrgphi2}) and (\ref{conformal_condition3}) we obtain
\be
\label{Tk2}
-\b^{d+2}\langle T_{\tau\tau}\rangle =-2(d+1)\cC_{d+2}(m\b)+2(m\b)^2a^{(d+2)}_{\cQ_{0;0}}\,.
\ee
Comparing (\ref{a_coeff4}) and (\ref{Tk2}) we see that the symmetric traceless tensor $\tilde{Q}_{\m\n}$ cannot just be the traceless part of the energy-momentum tensor $T_{\m\n}$, as in the unitary $k=1$ case, but it contains an additional part which we denote below as $\tilde{t}_{\m\n}$ 
\be
\label{QTt}
\tilde{Q}_{\m\n}=\left(T_{\m\n}-\frac{1}{d+2}\delta_{\m\n}T^\rho_{\,\,\rho}\right)+m^2\tilde{t}_{\m\n}\,.
\ee
Substituting (\ref{QTt}) into (\ref{a_coeff4}) and using (\ref{Tk2}) we find
\be
\label{QTt1}
-\frac{1}{d+2}\b^{d+2}\langle T^\rho_{\,\,\rho}\rangle+(m\b)^2\b^{d}\langle\tilde{t}_{\tau\tau}\rangle=\frac{2(m\beta)^2}{d+2}a^{(d+2)}_{\cQ_{0;0}}+\frac{d(d+1)(m\b)^2}{d+2}\left(a^{(d+2)}_{\cQ_{0;0}}+\frac{(m\b)^2}{2\pi d}a^{(d)}_{\cQ_{0;0}}\right)
\ee
We will now show that the symmetric traceless tensor $\tilde{t}_{\m\n}$ is the spin-2 pillar operator that gives rise to the ${\cal G}^{(d+2)}_{2;d}$ conformal block in (\ref{phiphi111}), as it is depicted in Fig. \ref{fig: conf integrals}. Indeed, the leading-twist coefficient of that block is read from (\ref{phiphi111}), (\ref{D_coeff}) and also (\ref{eq:attobt}) and (\ref{recursion}) to be
\be
\label{C2}
\frac{1}{4\pi}\cC^{(d+2,2)}_{2}=\frac{1}{2(d+2)}\left(a^{(d+2)}_{\cQ_{0;0}}+\frac{(m\b)^2}{2\pi d}a^{(d)}_{\cQ_{0;0}}\right)
=\frac{f_{\phi\phi\tilde{t}}}{c_{\tilde{t}}}\frac{2\beta^{d}\langle\tilde{t}_{\tau\tau}\rangle}{d(d+1)}\,,
\ee
where $f_{\phi\phi\tilde{t}}/c_{\tilde{t}}$ is the ratio of the coefficients of the relevant three- and two-point functions. Notice however that this ratio is not fixed by a Ward identity as $\tilde{t}_{\m\n}$ does not obey any conservation equation \cite{Brust:2016gjy}. Nevertheless for the special class of higher derivative Lagrangians (\ref{L_massive_scalar_1}) that we consider here, we can calculate the thermal trace of their energy-momentum tensor. This follows the observation that the free energy of (\ref{L_massive_scalar_1}) is $k$ times the free energy of the corresponding $k=1$ theory. Then following the reasoning in (\ref{Ttrace})-(\ref{traceT}) we obtain
\be
\label{Td+2}
\beta^{d+2}\langle T^\rho_{\,\,\rho}\rangle=-2(m\beta)^2a^{(d+2)}_{\cQ_{0;0}}=\frac{1}{\pi}(m\b)^2\cC_{d}(m\b)\,.
\ee
If we substitute (\ref{Td+2}) into (\ref{QTt1}) we find that (\ref{C2}) is satisfied if $f_{\phi\phi\tilde{t}}/c_{\tilde{t}}=1/4$. 

The first equality of (\ref{Td+2}) is simply the statement that the traces of the energy momentum tensors of the $k=1$ and $k=2$ theories in $d+2$-dimensions are proportional and hence their respective gap equations are the same.\footnote{This property holds for all theories with Lagrangians (\ref{L_massive_scalar_1}) in any fixed dimension and for all $k$.} 
The second equality of (\ref{Td+2}) shows that the thermal trace of the energy-momentum of the uplifted $d+2$-dimensional theory is proportional to the free energy of the original $d$-dimensional theory. From these we see that the gap equations of the original $d$-dimensional unitary theory and the uplifted $d+2$-dimensional ones are different. Hence if we apply our dimensional uplift to the nontrivial thermal two-point functions of the $d$-dimensional theory we will not obtain the nontrivial thermal two-point functions of the uplifted $d+2$-dimensional theory. In this sense the two underlying CFTs are not simply related by our dimensional uplift.  Furthermore, the operator content of the latter nonunitary CFT$_{d+2}$ is more complicated: we do not observe the dropping out of a whole class of operators, such as $\phi^2$ and its higher-twist column, as it happens in the original unitary CFT$_d$. Nevertheless, we find it interesting that we can associate the zero of the free energy of a unitary massive QFT$_d$ with a thermal trace of the energy-momentum tensor of a nonunitary CFT$_{d+2}$. We summarize below some of the salient features of the thermal dimensional uplift. 

\begin{itemize}[leftmargin=10pt]
\item
The higher-spin operators that give rise to the thermal conformal blocks in (\ref{phiphi111}) are split into two classes: I) would-be higher-spin conserved currents with dimensions $\Delta_{d+2,s}=d+s$ that saturate the unitarity bound in $d+2$-dimensions. II) higher-spin operators with dimensions $\Delta_{d+2,s}-2=d-2+s$ that fall below the unitarity bound in $d+2$-dimensions.  
\item
The contributions of the higher-twist operators corresponding to each of the two operator classes above are encoded into the dressing of the associated coefficients and they can be calculated from the unitary $k=1$ theory. 
\item The uplift introduces an operator mixing. One can see from (\ref{phiphi111}) that for a given spin the higher-twist contributions from the two classes I) and II) give rise to the same power of $r$ in the expansion when their twist orders differ by one. This applies also in the case of the energy-momentum tensor which mixes with a higher-twist  spin-2 symmetric traceless tensor. 
\item The energy-momentum contribution of the uplifted $d+2$-dimensional theory arises from the spin-4 tensor of the original $d$-dimensional theory. 
\item The dimensional uplift is valid for all values of the mass parameter $m$. Moreover the gap equation of the uplifted $d+2$-dimensional theory is equivalent to the vanishing of the free energy of the original $d$-dimensional theory and as a consequence the thermal gap equations of the original and the uplifted theories are different. This implies that the conformal thermal two-point functions of the original CFT$_d$ do not uplift to conformal thermal two-point functions of a higher derivative CFT$_{d+2}$ for $m\neq 0$. 
\end{itemize}

\section{Discussion and Outlook}\label{sec:discussion}

In this work we explored the idea that massive deformations of scalar thermal two-point functions can be used as diagnostic tools in the space of CFTs at zero temperatures. Such two-point functions can be expanded in thermal conformal blocks of would-be higher-spin currents with coefficients dressed by the resummed contributions of the higher-twist operators. The correlators are associated with a thermal CFT if the mass parameter $m$ satisfies a gap equation which expresses the vanishing of the thermal trace of the energy-momentum tensor. The nonzero solutions of the gap equation enable us to elucidate part of the structure and properties of nontrivial CFTs in general odd dimensions $d$. 

Our main result, however, is the dimensional uplift of the massive thermal two-point functions above. Specifically, we show that for all values of $m$ the thermal correlators of the unitary theory in $d$-dimensions can be expanded into $d+2$-dimensional thermal conformal blocks, and hence they can be interpreted as thermal two-point functions of higher dimensional theories. We identify the latter as a particular four-derivative, and hence nonunitary, massive theory.  The uplifted correlator can be associated with a CFT$_{d+2}$ when $m$ is such that the free energy of the original $d$-dimensional theory becomes zero. This sheds light on its operator structure and reveals a nontrivial operator mixing. Since the gap equations of the original unitary $d$-dimensional and the uplifted $d+2$-dimensional theories are not the same, we conclude that the corresponding conformal thermal two-point functions are not related by the dimensional uplift when the thermal mass is nonzero.  

Our work offers some new perspectives in the study of conformal thermal two-point functions and through them in the understanding of the associated CFTs. For a start it is hard not to suspect the relation of the thermal uplift with the famous Parisi-Sourlas uplift (see, e.g. \cite{Kaviraj:2019tbg,Trevisani:2024djr,Gliozzi:2025xmr} for recent works). For example our uplifted Lagrangians (\ref{L_massive_scalar_1}) should correspond to the bosonic part of the Parisi-Sourlas SUSY Lagrangian if we use the auxiliary field methods of \cite{Gibbons:2019lmj,Paci:2025pxo}. This could provide a clearer explanation for the association of the $d$-dimensional free energy of a unitary QFT to the thermal trace of a $d+2$-dimensional higher-derivative theory. Also, our higher-derivative theories (\ref{L_massive_scalar_1}) are a higher dimensional generalization of the degenerate limit of the Pais-Uhlenbeck oscillator \cite{Smilga:2008pr,Nicolis:2018lkg}. The latter is by itself an interesting system whose thermodynamics may be related to the thermodynamics of our higher-derivative theories. An upcoming work will touch upon both of these questions above. 

In another direction it would be nice to study the operator mixing induced by the thermal uplift in more complicated models. One approach would be to make the exponent $k$ continuous see, e.g. \cite{Benedetti:2023pbt}. Another would be to study the thermodynamics of diverse deformations of higher-derivative theories such as in \cite{Safari:2017tgs,Safari:2017irw,Safari:2021ocb,Stergiou:2022qqj,Guo:2023qtt,Herzog:2024zxm,Guo:2025edk,Guo:2026vmq}.  We have left untouched the whole chapter of the relations of the coefficients $a^{(d)}_{\cQ_{s,0}}$ with conformal integrals. In the works \cite{Petkou:2021zhg,Karydas:2023ufs,Karydas:2025tfs} it was shown that the dimension $d$ is mapped to the loop order $L$ of the conformal integrals as $d=2L+1$. It would be nice to understand how  the parameter $k$ maps into conformal integrals, and in particular to the generalized ones discussed in \cite{Giombi:2022gjj}. Finally, an obviously relevant question is the extension of the thermal uplift discussions for even $d$, which is expected to touch upon properties of two-dimensional CFTs and possibly also of string theory.

\acknowledgments

A.C.P. wishes to acknowledge discussions and correspondence with S. Giombi, E. Skvortsov, S. Solodukhin, A. Stergiou.  The work of S.L. was supported by  the Natural Science Foundation of China (Grant No. 12522504). We wish to acknowledge the use of freely available AI models in some of the calculations in the Appendix, and also Claude (Anthropic) for substantial help in digitising a hand-drawn version of Fig.1, and for checking a few misprints in the draft.

\appendix

\section{Notations}\label{app:notation}
The coordinates on the thermal geometry $S^1_\b\times\mathbb{R}^{d-1}$ are $x^\m=(\tau,\vec{x})$, $\tau\in[0,\beta]$. We also have $r^2=\tau^2+\vec{x}^2$, $\cos\theta=\tau/r$. The contribution of the unit operator $\mathbb{1}$ in the thermal two-point function (\ref{2ptCFTgen}) corresponds to the normalization of the corresponding zero temperature result. For the massive free scalars these are 
\be
\label{Cphi}
\cL_{d,k}=\frac{1}{2}\phi(-\Box+m^2)^k\phi \rightarrow \frac{1}{p^{2k}}\xrightarrow{F.T.}\frac{C_{\phi}(d,k)}{r^{d-k}}\,,\,\,\,C_{\phi}(d,k)=\frac{\Gamma\left(\frac{d}{2}-k\right)}{4^k\pi^{\frac{d}{2}}\Gamma(k)}\,.
\ee
The normalization constant of the two-point function is related to the solid angle in $d$-dimensions $S_d=2\pi^{d/2}/\Gamma(d/2)$ as $(d-2)S_dC_\phi(d)=1$ for $d>2$.

\section{The expansion of the zero temperature massive two-point function}\label{app:expansionof0T}
In this section we will compute the coefficients in the expansion in \eqref{zeroT}. Defining $x=mr$, we can rewrite it  as
\begin{equation}
    \frac{1}{(2\pi)^{\frac{d}{2}}}x^\nu K_\nu(x) = \sum_{k=0}^{\infty}A_{2k}x^{2k}+\sum_{k=0}B_{2k+1}x^{2k+1}\,,
\end{equation}
where $A_{2k}$ and $B_{2k+1}$ are related to the $A_{2k}^{(d)}(m\beta)$ and $B_{2k+1}^{(d)}(m\beta)$ in \eqref{zeroT} by
\begin{equation}
    A_{2k}^{(d)}(m\beta) = A_{2k}(\beta m)^{2k}\,,\, B_{2k+1}^{(d)}(m\beta) = B_{2k+1}(\beta m)^{2k+1}\,.
\end{equation}
Starting from the definition of $K_\nu(x)$ we know that
\begin{equation}
    K_\nu(x) = \frac{\pi}{2}\frac{I_{-\nu}(x)-I_{\nu}(x)}{\sin\left(\pi\nu\right)} = (-)^{\nu-\frac{1}{2}}\frac{\pi}{2}\left(I_{-\nu}(x)-I_{\nu}(x)\right)\,,
\end{equation}
where we have used in the second equality that in odd dimensions $\nu=\frac{d}{2}-1$ is a half odd number. $I_\alpha(x)$ is the modified Bessel function of the first kind defined by
\begin{equation}
    I_\alpha(x) = \sum_{m=0}^{\infty}\frac{1}{m!\Gamma(m+\alpha+1)}\left(\frac{x}{2}\right)^{2m+\alpha}\,.
\end{equation}
We can then find
\begin{equation}
    x^{\nu}K_{\nu}(x) = (-)^{\nu-\frac{1}{2}}\frac{\pi}{2}\left[\sum_{m=0}^{\infty}\frac{1}{2^{2m-\nu}m!\Gamma\left(m-\nu+1\right)}x^{2m} - \sum_{m=0}^{\infty}\frac{1}{2^{2m+\nu}m!\Gamma\left(m+\nu+1\right)}x^{2m+2\nu}\right]\,.
\end{equation}
Since $2\nu=d-2$ is odd, we can immediately see that the even powers get contribution only from the first term, while the odd powers get contribution only from the second term. One can then read that
\begin{equation}
    A_{2k} = \frac{(-)^{\nu-\frac{1}{2}}}{(2\pi)^{\frac{d}{2}}}\frac{\pi}{2}\frac{1}{k!2^{2k-\nu}\Gamma\left(k-\nu+1\right)}\,,
\end{equation}
$B_{2k+1}$ vanishes for $k<\nu-\frac{1}{2}$ and
\begin{equation}
    B_{2k+1} = \frac{(-)^{\nu+\frac{1}{2}}}{(2\pi)^{\frac{d}{2}}}\frac{\pi}{2} \frac{1}{(k-\nu+\frac{1}{2})!2^{2k+1-\nu}\Gamma(k+\frac{3}{2})} = \frac{(-)^{\frac{d-1}{2}}}{2^{k+2}\pi^{\frac{d-1}{2}}\left(k-\frac{d-3}{2}\right)!(2k+1)!!}
\end{equation}
for $k\geqslant\nu-\frac{1}{2}$. We can also write the $A_{2k}$'s using double factorials. We need to do it for two different cases. When $k<\nu-\frac{1}{2}$, using
\begin{equation}
    \Gamma(k-\nu+1) = \frac{\pi}{\sin\left(\nu\pi-k\pi\right)}\frac{1}{\Gamma\left(\nu-k\right)} = \frac{(-)^{\nu-k-\frac{1}{2}}2^{\nu-k-\frac{1}{2}}\sqrt{\pi}}{(2\nu-2k-2)!!}\,,
\end{equation}
we find
\begin{equation}
    A_{2k} = \frac{(-)^k(d-4-2k)!!}{(2\pi)^{\frac{d-1}{2}}2^{k+1}k!}\,.\label{Aksmall}
\end{equation}
For the case where $k\geqslant\nu-\frac{1}{2}$, one can directly use the relation between Gamma functions and double factorials to find
\begin{equation}
    A_{2k} = \frac{(-)^{\frac{d-3}{2}}}{(2\pi)^{\frac{d-1}{2}}2^{k+1}k!(2k-d+2)!!}\,.\label{Aklarge}
\end{equation}
Note that when $k=\nu-\frac{1}{2}$,  \eqref{Aksmall} and \eqref{Aklarge} give the same result and thus we can use either form in this case.

The even-power coefficients $A_{2k}^{(d)}$ give rise to singular contributions as $r\rightarrow 0$ for  $0 \le k \le \nu - 1/2$ and regular ones  for  $k \ge \nu + 1/2$
\begin{equation}
A_{2k}^{(d)}(m\beta) = 
\begin{cases} 
\displaystyle (-1)^k \; \frac{(d - 2k - 4)!! \,(m\beta)^{2k}}{2^{k+1} (2\pi)^{\frac{d-1}{2}} \, k!} \, & \text{for } 0 \le k \le \nu - \frac{1}{2} \\[14pt]
\displaystyle (-1)^{\frac{d-3}{2}} \; \frac{(m\beta)^{2k}}{2^{k+1} (2\pi)^{\frac{d-1}{2}} \, k! \, (2k - d +2)!!}
, & \text{for } k \ge \nu + \frac{1}{2} 
\end{cases}
\end{equation}
where we use the definition$(-1)!!=1$.
The odd-power coefficients $B_{2k+1}^{(d)}$ vanish identically for $k < \nu - 1/2$ and are given by
\begin{equation}
B_{2k+1}^{(d)}(m\beta) = 
\begin{cases}
0, & \text{for } 0 \le k < \nu - \frac{1}{2} \\
\displaystyle (-1)^{\frac{d-1}{2}} \frac{1}{2^{k+2} \pi^{\frac{d-1}{2}}} \frac{(m\beta)^{2k+1}}{(2k+1)!! \, (k-\frac{d-3}{2})!}, & \text{for } k \ge \nu - \frac{1}{2}
\end{cases}
\end{equation}
Notice that the singular contributions alternate sign as $(-1)^k$, while for all other contributions the sign is determined overall by the dimension $d$.  Two notable terms are the coefficient of the $m$-independent term $A_0^{(d)}$ and the $r$-independent contribution $B_{d-2}^{(d)}(m\b)$ given respectively as
\be
\label{A0}
A_0^{(d)}=\frac{\Gamma\left(\frac{d}{2}-1\right)}{4\pi^{\frac{d}{2}}}\equiv C_\phi(d,1)\,,\,\,\,\,B_{d-2}^{(d)}(m\b)=(-1)^{\frac{d-1}{2}}\frac{(m\b)^{d-2}}{2(2\pi)^{\frac{d-1}{2}}(d-2)!!}\,.
\ee

\section{Spatial coincident limit and  correspondence with the harmonic oscillator for $d=1$.}\label{app:d=1}

At the spatial coincident limit $\cos\theta = 1$, $\vec{x}=0$ the Gegenbauer polynomials simplify to
\begin{equation}
C_s^{d/2-1}(1) = \binom{s+d-3}{s} = \frac{(s+d-3)!}{s!\,(d-3)!}\,.
\end{equation}
Substituting into $g_{d,1}(r, \vec{x})$ yields
\begin{align}
\label{gd10}
g_{d,1}(\tau,\vec{0}) &= \frac{1}{\tau^{d-2}}\sum_{k=0}^{\infty}A_{2k}^{(d)}\left(\frac{\tau}{\beta}\right)^{2k} + \frac{1}{\beta^{d-2}}\sum_{s=0,2,4,\dots}\binom{s+d-3}{s} \mathcal{B}^{(d)}_{s}(m\beta, m\tau) \left(\frac{\tau}{\beta}\right)^s \nonumber\\
\end{align}
This general result has an interesting $d=1$ limit.\footnote{It should be noted that this is not a limit in the strict mathematical sense as $d$ takes discrete values.} We first notice that the combinatorial prefactor vanishes for all even spin modes $s \ge 2$ when $d \to 1$
\begin{equation}
\lim_{d \to 1} \frac{\Gamma(s+d-2)}{s! \, \Gamma(d-2)} = \delta_{s, 0}\,,
\end{equation}
reflecting the fact that there is no spatial angular momentum in one dimension. Hence, only the $a^{(1)}_{\cQ_{0;0}}$ term would contribute in the second sum of (\ref{gd10}). However such a term does not follow from our results (\ref{aQs0}) which are valid for $d\geq 3$. Instead, we can use the physical interpretation of this coefficient as being the thermal average $\langle\phi^2\rangle$. In $d=1$ we can identify this with $\langle \hat{x}^2\rangle$ where $\hat{x}$ is the position operator of a harmonic oscillator with unit mass and frequency denoted by $m$. Hence we define
\be
\label{a100}
a^{(1)}_{\cQ_{0;0}}\equiv \langle \hat{x}^2\rangle =\frac{1}{2m}\left(1+2n(m\b)\right)\,,
\ee
with $n(m\b)= 1/(e^{m\b}-1)$ the thermal Bose factor. Using $I_{-1/2}(m\tau) = \sqrt{\frac{2}{\pi m\tau}} \cosh(m\tau)$ and 
\begin{equation}
r \sum_{k=0}^{\infty} A_{2k}^{(1)}\left(\frac{\tau}{\beta}\right)^{2k} = - \frac{\tau}{2} \sum_{k=0}^{\infty} \frac{(m\tau)^{2k}}{(2k+1)!} = -\frac{\sinh(m\tau)}{2m}
\end{equation}
we find
\begin{equation}
\label{g11}
g_{1,1}(\tau, \vec{0})=\langle \hat{x}(\tau)\hat{x}(0)\rangle = \frac{e^{-m\tau}}{2m} + \frac{\cosh(m\tau)}{m\left(e^{m\beta} - 1\right)}\,,
\end{equation}
which is the well-known thermal two-point function\footnote{There are many ways to obtain this. One is to make a Wick rotation and then calculate the thermal average of $\hat{x}(\tau)\hat{x}(0)=e^{\tau \hat{H}}\hat{x}(0)e^{-\tau\hat{H}}\hat{x}(0)$.} of the harmonic oscillator with unit mass and frequency $m$. In the spirit of the main body of this work we might define a gap equation for this system as
\be
\label{gapho}
a^{(1)}_{\cQ_{0;0}}=0\Rightarrow m\mapsto m_p=i\frac{\pi}{\beta}(2p+1)\,,\,\,\,p\in \mathbb{Z}\,.
\ee
Substituting $m_0$ into the two-point function (\ref{g11}) we find
\be
\label{g11cofn}
g_{1,1}(\tau)=-\frac{\beta}{2\pi}\sin\left(\frac{\pi\tau}{\beta}\right)\,.
\ee
On the one hand this may be identified with a two-point function of an operator with dimension $\Delta=-1/2$ in a one-dimensional thermal CFT. On the other hand, following our thermal uplift suggestion this could be identified with the thermal two-point function of a four-derivative CFT in three-dimensions at spatially coincident points. In any case one is tempted to observe similarities of the above discussion with recent work on the imaginary distance bound \cite{Maldacena:2026jqd} to which we hope to return soon.

\section{The product expansion for classical massive free scalars}\label{app:productexpansion}

At the classical field theory level one can analytically calculate the product of two scalar fields into an expansion of symmetric traceless representations of the rotation group $SO(d)$ using classical $d$-dimensional Rayleigh plane-wave expansion.
We begin by substituting the formal expression 
\begin{equation}
    \phi(x) = e^{x \cdot \partial} \phi(0)\,,
\end{equation}
into the symmetric product
\begin{equation}
    \mathcal{O}_{\text{sym}}(x) \equiv \frac{1}{2} \Big[ \phi(x)\phi(0) + \phi(-x)\phi(0) \Big]\,.
\end{equation}
 The  Taylor expansion of the field translation operator is given by
\begin{align}
\frac{1}{2}\left(e^{x \cdot \partial} +e^{-x\cdot\partial}\right)&= \sum_{n=0}^{\infty} \frac{1}{n!} x^{\mu_1} \dots x^{\mu_n} \partial_{\mu_1} \dots \partial_{\mu_n}\nonumber \\
&= \sum_{s \text{ even}} \sum_{t=0}^{\infty} C(s,t) (x^2)^t x^{\mu_1} \dots x^{\mu_s} D_{\mu_1 \dots \mu_s}^{(s)} \Box^t\,.
\end{align}
The irreducible symmetric traceless rank-$s$ tensor operator $D^{(s)}_{\mu_1\cdots\mu_s}$ can be explicitly constructed from  partial derivatives via a nested series of metric trace subtractions in $d$ dimensions as
\begin{equation}
D_{\mu_1 \dots \mu_s}^{(s)} = \partial_{\mu_1}\dots\partial_{\mu_s} - \frac{1}{d+2s-4} \sum_{\text{dist.}} \eta_{\mu_1\mu_2}\partial_{\mu_3}\dots\partial_{\mu_s}\Box + \dots
\end{equation}
Inverting this trace-removal formula allows us to express a string of partial derivatives as a linear combination of traceless operators decorated with flat metrics $\eta_{\m\n}$ 
\begin{equation}\label{inversion}
\partial_{\mu_1}\dots\partial_{\mu_{2m}} = \sum_{t} \left[ \prod_{j=1}^t \frac{1}{d+2s+2j-2} \right] \sum_{\text{dist.}} \eta_{\mu_1\mu_2}\dots\eta_{\mu_{2t-1}\mu_{2t}} D_{\mu_{2t+1}\dots\mu_{2m}}^{(s)} \Box^t
\end{equation}
When we contract this identity against the outer coordinate string $x^{\mu_1}\dots x^{\mu_{2m}}$, each metric $\eta_{\mu_i\mu_j}$ absorbs two coordinate indices to generate a scalar length factor $x^2$, producing a cumulative scaling term $(x^2)^t$.

To identify the overall multiplier for a given $(s,t)$ subspace, we count the exact number of distinct ways to partition $2m$ total coordinate indices into $t$ pairs (which condense into $(x^2)^t$), leaving exactly $s$ indices uncontracted to match $D^{(s)}$. This counting is governed by standard combinatorics:
\begin{equation}
N = \binom{2m}{2t} \times (2t-1)!! = \frac{(2m)!}{(2t)! s!} \times \frac{(2t)!}{2^t t!} = \frac{(2m)!}{2^t t! s!}
\end{equation}
Substituting this combinatorial coefficient back into the raw Taylor coefficient $\frac{1}{(2m)!}$ yields
\begin{equation}
\frac{1}{(2m)!} \left[ \frac{(2m)!}{2^t t! s!} \prod_{j=1}^t \frac{1}{d+2s+2j-2} \right] (x^2)^t x^{\mu_1}\dots x^{\mu_s} D_{\mu_1\dots\mu_s}^{(s)} \Box^t
\end{equation}
The factorial $(2m)!$ cancels out perfectly. This yields the universal master formula for the trace descendant coefficient
\begin{equation}\label{master}
C(s,t) = \frac{1}{2^t \cdot t! \cdot s! \cdot \prod_{j=1}^t (d + 2s + 2j - 2)}
\end{equation}
We rewrite the finite product chain using the standard Pochhammer symbol 
\begin{equation}
\prod_{j=1}^t (d + 2s + 2j - 2) = 2^t \prod_{j=1}^t \left( \frac{d}{2} + s - 1 + j \right) = 2^t \left( \frac{d}{2} + s \right)_t
\end{equation}
The Laplacian operator $\Box \equiv \partial^\alpha \partial_\alpha$ commutes  with the symmetric traceless spatial derivatives $D^{(s)}$ and we can isolate the infinite summation over the trace descendant index $t$:
\begin{equation}
    \sum_{t=0}^{\infty} \frac{1}{t! \left(\frac{d}{2} + s\right)_t} \left( \frac{x^2 \Box}{4} \right)^t
\end{equation}
This series matches the exact definition of the confluent hypergeometric function ${}_0F_1(b; z) = \sum_{t=0}^{\infty} \frac{z^t}{t! (b)_t}$ and using the  e.o.m. $\Box\phi =m^2\phi$ we obtain
\begin{equation}
\label{Osymm}
    \mathcal{O}_{\text{sym}}(x) = \sum_{s=0, 2, 4 \dots}^{\infty} \frac{x^{\mu_1}\dots x^{\mu_s}}{s!} \, {}_0F_1\left(\frac{d}{2}+s; \, \frac{x^2 m^2}{4}\right) \phi(0) D_{\mu_1\dots\mu_s}^{(s)}\phi(0) 
\end{equation}
For each spin, the operators on the r.h.s. contain the resummation of all higher-twist contributions. We can identify the higher-spin operators as 
\be
\label{HSoperators}
\cQ_{s;t}=-\phi(0) D_{\mu_1\dots\mu_s}^{(s)}\Box^t\phi(0) \,,
\ee
to be consistent with the Ward identities of \cite{Petkou:1994ad}. For $t=0$ these are the higher-spin conserved currents and when we take the thermal average of (\ref{Osymm}) using (\ref{eq:1ptfunction}) and the identity
\be
\label{Gegen}
x^{\mu_1}\dots x^{\mu_s}(e_{\m_1}\dots e_{\mu_s}-\text{traces})=\frac{s!}{2^s(\n)_s}\,,
\ee
we obtain the coefficients (\ref{eq:atob}) with $f_{\phi\phi\cQ_s}/c_{\cQ_s} =-1/s!$ in agreement with the result in \cite{Skvortsov:2015pea,Benedetti:2023pbt}.

\bibliographystyle{JHEP}

\providecommand{\href}[2]{#2}\begingroup\raggedright\endgroup


\end{document}